\documentclass[10pt,journal,compsoc]{IEEEtran}
\ifCLASSOPTIONcompsoc
  \usepackage[nocompress]{cite}
\else
                                                                                                                                                            \usepackage{cite}
\fi
\ifCLASSINFOpdf
\else
\fi

\usepackage[pdftex]{graphicx}
\usepackage{float}
\usepackage{color,soul}
\restylefloat{figure}
\begin{document}

\title{Classifying discourse in a CSCL platform \\to evaluate correlations with \\Teacher Participation and Progress}

\author{Eliana~Scheihing,~\IEEEmembership{}
Matthieu~Vernier,~\IEEEmembership{}
Javiera~Born,~\IEEEmembership{}
Julio~Guerra,~\IEEEmembership{}
Luis C\'arcamo~\IEEEmembership{}
}

\IEEEtitleabstractindextext{%
\begin{abstract}
In Computer-Supported learning, monitoring and engaging a group of learners is a complex task for teachers, especially when learners are working collaboratively: \textit{Are my students motivated? What kind of progress are they making? Should I intervene? Is my communication and the didactic design adapted to my students? } Our hypothesis is that the analysis of natural language interactions between students, and between students and teachers, provide very valuable information and could be used to produce qualitative indicators to help teachers' decisions. We develop an automatic approach in three steps (1) to explore the discursive functions of messages in a CSCL platform, (2) to classify the messages automatically and (3) to evaluate correlations between discursive attitudes and other variables linked to the learning activity. Results tend to show that some types of discourse are correlated with a notion of Progress on the learning activities and the importance of emotive participation from the Teacher.

\end{abstract}
\begin{IEEEkeywords}
Collaborative learning tools, Consequences for learning, Discourse, Machine Learning.
\end{IEEEkeywords}}

\maketitle

\IEEEdisplaynontitleabstractindextext

\IEEEpeerreviewmaketitle

\IEEEraisesectionheading{\section{Introduction}\label{sec:introduction}}

\subsection{ICT challenges in Education}

The emergence of massive open online courses (\textit{Coursera, edX, OpenClassrooms}), e-learning platforms (\textit{Moodle, Khan-Academy}) and Computer-Supported Collaborative Learning platforms (CSCL) (\textit{Metafora, Claroline}) offers a new horizon in terms of knowledge building and sharing. At the same time, Blended Learning \cite{2014-BlendedLearning-Guzer-1} is a growing pedagogical program in formal education combining traditional classroom sessions and distance learning via web platforms such as the \textit{Kelluwen project} \cite{ref:scheihing2013}. Several reflections accompany this paradigm shift in Education. We regroup these lines of research under four challenges: 
\begin{enumerate}[\IEEEsetlabelwidth{10}]
\item how to optimize the motivation and satisfaction of learners ? (also called \emph{engagement})
\item how to optimize the skills acquisition and the performance of learners ? (also called \emph{progress})
\item how to predict learning behaviours (engagement / discouragement, progress / mental block, learning style, etc.) ?
\item how to optimize the role of the teacher ?
\end{enumerate}
To address these challenges, existing IT research is experimenting and evaluating several ideas:
\begin{enumerate}[\IEEEsetlabelwidth{10}]
\item gamifying learning experiments \cite{2013-CE-Dominguez-1} \cite{2014-TLT-Ibanez-1} \cite{2016-CE-Hooshyar-1},
\item personalizing learning experiments according to learner profiles \cite{2014-TLT-Yang-1}\cite{2014-CE-Feldman-1} \cite{2015-BJET-Rahimi-1},
\item improving the portability of learning experiments on mobile devices and connected objects to facilitate activities out of classroom \cite{2015-TLT-ChinLeeChen-1}\cite{2015-TLT-LooiSunXie-1},
\item adding visual feedbacks to increase the learners awareness of their behaviours \cite{2015-TLT-AuvinenHakulinenMalmi-1}\cite{2015-TLT-DemmansBull-1}\cite{brusilovsky2015value},
\item developing interactions and collaborative learning between learners \cite{2014-CSCL-Engelmann-1}, \cite{2016-CE-Goggins-1}, \cite{2015-CE-Vanderhoven-1}.
\end{enumerate}

In the majority of these researches, learners' progress and engagement are evaluated on the basis of structured data available in the platform: test results, time spent on the platform, number of activities realized, number of messages posted or awards received, etc. 
These indicators give valuable quantitative approximations of progress and engagement levels but are limited to provide a more qualitative understanding of these cognitive phenomena.

\subsection{The Natural Language: A source of data to explore the cognitive phenomena related to Learning}

In the context of CSCL, special attention is given to the development of interactions between learners. In the collaboration theory, it is assumed that knowledge is not only a question of memorising a list of facts, but is constructed dynamically through social relationships, shared activities and dialogues\cite{2004-Book-Stahl-1}. For this reason, social media (forums, chat rooms, microblogs, video-conferences) are particularly central in CSCL platforms \cite{2007-CHB-Janssen-1} and socio-communicative skills play a key role in knowledge building.

By analysing dialogues between learners in CSCL platforms, previous studies have shown that various linguistic and cognitive phenomena influence or are correlated with progress and engagement of learners:
\begin{enumerate}
    \item mutual listening and dialogical cohesion \cite{2014-CSCL-Wise-1} \cite{2015-CSCL-Dascalu-1},
    \item expression of positive/negative emotions  \cite{2015-CE-Yeha-1} \cite{2014-TLT-Chen-1},
    \item expression of uncertainty and questioning \cite{2014-CSCL-Jordan-1} \cite{2011-ALT-Zhang-1},
    \item quality of argumentation and vocabulary used \cite{2013-CSCL-Alagoz-1} \cite{2010-AI-Dascalu-1} \cite{2014-TLT-Jain-1},
    \item expression of agreement/disagreement and conflict resolution \cite{2013-CSCL-Weinberger-1},
    \item expression of leadership and work organization between learners \cite{2014-CSCL-Mercier-1},
    \item teacher's active participation in discussions \cite{2014-TLT-Brinton-1}.
\end{enumerate}

Some of these studies have adopted an automated approach, using Natural Language Processing and Machine Learning techniques \cite{2014-Dascalu-1}. However, significant efforts are still required, on the one hand, to automate textual analysis of complex discursive phenomena in the context of Education (expression of opinion, doubt, belief, disagreement, arguments, etc.), and on the other hand, to evaluate how this new data could be used to produce relevant qualitative indicators and provide assistance in teachers' decision making.

\subsection{Overview of the research problem addressed in this article}

In this article we automate the analysis of the learners' and teachers' discourse in a CSCL platform and then study how discursive phenomena are correlated with other variables such as the quantitative participation of teachers, the level of progress of the learning activities, or the didactic design used. We believe that our work can be used to better support teachers in monitoring the development of learning activities, as well as support instructional designers in developing better learning activities.  

Our methodology is divided in three steps. First, we use machine learning techniques to explore latent topics and discover discursive categories present in the messages from learners of teachers. Secondly, we automatically classify the messages according to their main discursive category. Finally, we explore correlations between different variables related to the learning activities and the discursive behaviours observed in the previous step.

\begin{figure}
\begin{center}
\includegraphics[trim={0 150pt 0 0},clip,width = .49\textwidth]
{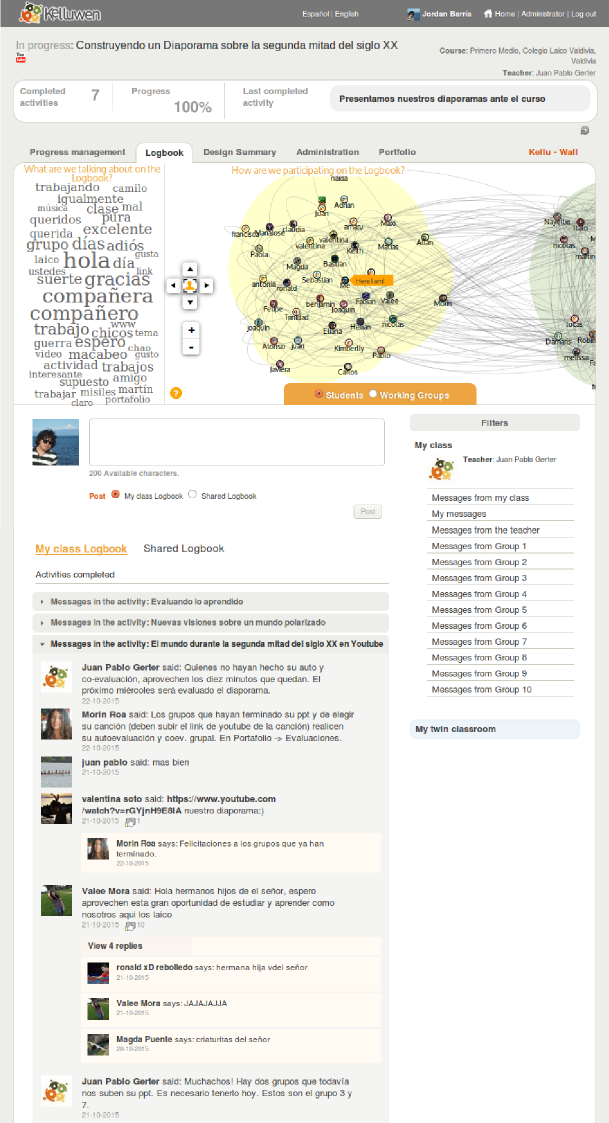}
\caption{Overview of the microblogging tool (the "Logbook") in the Kelluwen web platform during a collaborative learning activity in a chilean middle school : \emph{Building a slideshow about the second half of the 20th century}.}
\label{figKelluwen}
\vspace{-1.0em}
\end{center}
\end{figure}

The data used in our experiments was taken from the CSCL platform Kelluwen \cite{ref:scheihing2013}. Since 2011, Kelluwen web platform is used in formal curriculum of public education in southern Chile by about 5000 students in 182 classes of middle school and high school. It includes a microblogging tool called the Logbook (Figure \ref{figKelluwen}) where students and teachers can interact with classmates or other students following the same activity in other schools \cite{ref:inalef2012}.

The paper is organized as follows. The next section briefly presents useful work for our methodology in the fields of Natural Language Processing (NLP), Machine Learning (ML) and Linguistics. Section 3 describes the Kelluwen platform and dataset on which our experiments are based. In the section 4, 5 and 6, we explain the three steps of our methodology and precise intermediate results for each step. Finally, section 7 summarises our conclusions.

\section{Related work}

\subsection{Related work in NLP}

In Natural Language Processing (NLP), Latent Dirichlet allocation (LDA) \cite{2003-MLR-Blei-1} is a topic modeling method often used for identifying thematics from a large volume of unlabeled documents. The general assumption in LDA is that each document is generated from a latent set of probable topics, and each topic is a probability distribution among the entire vocabulary of the collection. LDA analyses co-occurrence of words in the documents to generate a set of topics, each of them as a distribution of probabilities over the vocabulary, by following two optimization goals: minimize the number of probable topics related to each document (describe each document as concisely as possible) and minimize the number of highly probable words within each topic (makes topics as specific as possible). By grouping words that occur together, LDA has the advantage to deal with synonyms and polysemy, which is of great use in our case where we have a considerable amount of miswritten words and use of slang.

Support Vector Machines (SVM) are a set of supervised machine learning techniques that has been widely and successfully used for text classification \cite{2002-SVM-Joachims-1}. SVM were originally designed for binary classification, but it is also able to process multiclass tasks by constructing a set of binary classifiers and using a strategy "one versus the rest" or "one versus one". During the learning phase, from a sample of training examples (for which the category is known), SVM builds a representation of the examples as points in a high-dimensional space. Then, it calculates a mathematical function, the SVM model, separating and optimizing the margin between the points of both classes.
During the test phase, new examples are mapped into that same space and predicted to belong to a category based on which side of the model they fall on.
SVM models have been particularly experienced and evaluated in the context of subjectivity analysis and opinion mining to decide if a text is subjective/objective \cite{2008-Subjectivity-Wilson-1} or if it is positive/negative/neutral \cite{2011-SVM-Saleh-1}. 
In this kind of issues dealing with the discursive level of language, SVM is interesting because it enables to manage a high number of dimensions of different natures. 

Finally, approaches combining LDA and SVM have been explored in several fields for classifying the stylometry (linguistic style) \cite{2009-Arun-1} or oral transcripts \cite{2014-LREC-Morchid-1} for example. This strategy has the interest of not presupposing any specific categories while taking advantage of SVM. In the particular context of Education, the recent work of Dascalu \cite{2014-Dascalu-1} presents promising results to automate discourse analysis with NLP and ML. His work focus on textual cohesion in dialogues between learners and shows that mutual listening and vocabulary sharing are qualitative indicators of engagement and progress in CSCL. 

\subsection{Related works in Linguistics}

\begin{figure}
\begin{center}
\includegraphics[width = .45\textwidth]
{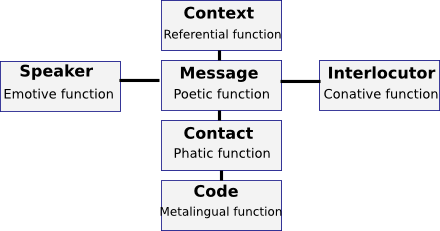}
\caption{Jakobson's semiotic theory model describing the 6 main functions of language in verbal communication.}
\label{jakobson-scheme}
\vspace{-1.0em}
\end{center}
\end{figure}

In Linguistics, the concept of Discourse implies that the true meaning of a sentence can't be assigned by its only linguistic construction but is largely determined by the communication situation within which it was enounced. In other words, to properly understand a sentence, a first step is to figure out what its main function is in the communication situation and to answer the question, '\textit{With what intention is this message transmitted?}'.
The well-known model of the functions of language introduced by Roman Jakobson \cite{1973-Jakobson-1} (Figure~\ref{jakobson-scheme}) defined six main communication functions, according to which an effective act of verbal communication can be described:

\begin{enumerate}[\IEEEsetlabelwidth{10}]
\item the \emph{emotive} function: the intention of the message is focused on the emotions and attitudes of the speaker towards what he is talking about.
Example: \textit{"I have trouble understanding the exercise"}
\item the \emph{conative} function: it puts the emphasis on the interlocutor seeking to influence his behaviour by an order or by persuasion. Example: \textit{"Move into groups of two or three"}
\item the \emph{phatic} function:  the intention is focused on creating a communicative channel, verifying that the channel works properly or simply drawing attention. Example: \emph{Hello! Can you hear me ?}
\item the \emph{metalingual} function: it aims to verify that the interlocutors understand each other, speak the same code or agree on the meaning of a particular word. Ex.: \textit{"What do you mean by 'statistics'?"}
\item the \emph{referential} function: it aims to give an information about the interlocutors context. Ex.: \textit{"The whale is a mammal"}
\item the \emph{poetic} function: it brings to light the message itself, the choice of certain words, their order, their sound, etc. This function aims to express the pleasure and the art of communicating. 
\end{enumerate}
Later, other models have been developed to specify subcategories for each of these functions. For example, Charaudeau \cite{1992-Charaudeau-1} distinguishes several categories in the emotive function such as expression of a "belief", an "agreement/disagreement", a "judgement", etc. 
 \medbreak
In the context of Education, little scientific research has been published to classify messages according to their discursive functions. Our hypothesis is that this kind of analysis could provide qualitative information: (1) to measure, or even predict, learner's progress and engagement, (2) to improve the teacher awareness on the development of the activity and (3) to evaluate the effects produced by teachers' communication.

\section{The Kelluwen project and the dataset}

The project Kelluwen is an attempt to incorporate the effective use of Blended Learning in formal curriculum of public education in southern Chile. It specifically targets the development of socio-communicative skills of middle and high school students and the use of a computer-supported collaborative tools as didactic resources.
It includes a CSCL web platform\footnote{Available under free license at http://app.kelluwen.cl} designed to support the learning activities by providing access to material and tools to facilitate the communication and monitoring of the work (Fig.~\ref{figKelluwen}).

\subsection{The Kelluwen Web platform}

In the project, a series of \emph{Didactic Designs} (DD) has been developed to work on three subjects: Language, History and Technology. A DD covers a unit of content as specified in the Chilean public school curriculum, and contains a series of activities which blend traditional classroom learning strategies with the use of social web tools like Wordpress, Youtube, etc. In this way, Kelluwen aims to bring the sharing and communication capabilities of the social web to formal learning activities in schools.

A typical DD contains a sequence of 8 to 12 classroom activities of 1 and 1/2 hours of duration (typical lecture time). In the DD, the students work in teams with the goal of creating an outcome that is finally published in a social Web tool, enabling further peer review processes. To make this happen, activities are organized into three stages: 
\begin{enumerate}
    \item a \emph{motivation stage}, where students are introduced to concepts and topics, and they must study the topic and begin to incorporate the knowledge;
    \item a \emph{creation stage}, where using the learned concepts they must generate new content, such as a blog or a video, exercising their communicative skills;
    \item and an \emph{evaluation stage}, where the generated content is now passed to other students, which may be from the same class group or from a different class group (even from a different school) working in the same DD, who must evaluate and give feedback on the work done. They also have the opportunity to do an auto-evaluation and receive comments from the guiding teacher. 
\end{enumerate}

During the activities, students and teachers have access to the Kelluwen platform. Although the main interaction between students and teachers happens in person, students and teachers tend to largely use it in various ways.


\subsection{The dataset}

We collected microblogging posts of 110 class groups working on 31 different Didactic Designs (DD) between 2011 and 2015 (during this period, the Logbook use was optional). Finally, the dataset contains 21.212 messages written in Chilean Spanish. The average length of a message is 7.3 words and 44.8 characters, making it quite short even for a microblog post. Additionally, since the logbook was explicitly intended as an informal channel, the messages contains heavy use of slang, emoticons and miswritten words. The messages are distributed according to the following criteria:
\begin{itemize}
    \item 20128 messages from students, 1084 from teachers ;
    \item 9631 from middle school, 11581 from high school ;
    \item 11341 messages expressed during 51 didactic experiences in Language, 8226 in History (43 experiences), 1655 in Technology (17 experiences).
\end{itemize}

Students' messages distribution in class groups has a median of 135, mean of 185, a standard deviation of 180, and a maximum of 1104 messages in one of the class groups. Teachers' distribution has a median of 5, mean of 10, and a maximum of 56 messages in one of the groups.


\subsection{Data preprocessing}


In order to facilitate the machine learning processes presented in the sections 4 and 5, we developed a series of natural language preprocessing to normalize the representation of the messages. 
\begin{enumerate}
    \item \textit{Tokenization}: a message is segmented into words ;
    \item \textit{Lemmatization}: the words are annotated with their generic forms (the lemma). For example, the lemma of \textit{actividades} (activities) is \textit{actividad} (activity). For this module, we used the lemmatization model for spanish available in the TreeTagger tool \cite{1995-TTG-Schmid-1},
    \item \textit{Filtering of stop-words}: a short list of grammatical words (such as \textit{aquel} (that), \textit{cada} (each)) are filtered to simplify the vectorial representation of messages,
    \item \textit{Basic spell correction}: in the same idea, several slight misspellings are corrected using regular expressions. In particular, it concerns some words containing several vowels running such as \textit{buuenooooo} (goooood) which are corrected to \textit{bueno} (good),
    \item \textit{Emoticons detection}: emoticons (;-), \textless3, :D, etc.) are detected with regular expressions and are considered like normal words,
    \item \textit{URL generalization}: all URL links are substituted by the word \textit{http}.    
\end{enumerate}
All these preprocessing modules\footnote{The dataset and preprocessing modules are availables at the following address under free license: https://github.com/matthieuvernier/.
} have been developped within the Apache framework UIMA\footnote{http://uima.apache.org/} (Unstructured Information Management Architecture).

\section{Exploring topics in  messages}

The first step of our methodology is mainly exploratory. We adopt a LDA non-supervised approach to uncover latent topics in the dataset. The goal is to obtain a first insight on the content of learners' and teachers' messages to help us understand the linguistic attitudes on the Kelluwen CSCL platform.

\subsection{Topic extraction with LDA}

LDA requires to postulate the number of clusters (or topics) as a parameter. After applying the preprocessing described in the section 3.3, we ran the topic modeling for 3, 4, 5 and 6 clusters using \textit{R} and the package \textit{lda} with the Gibbs sampling algorithm\footnote{parameters used: burn-in=100
000 and iterations=500.000.}. We repeat each experiment 3 times and compute log-likelihood. 

Results of maximum log-likelihood run for the experiment with 3 topics are shown in the table~\ref{topwords} where the Top-20 most probable words within each topic are shown. A review of the probabilities within each topic reveals that the Top-20 words cover 34\% (for Topics~2 and~3) and 39\% (for Topic~1) of the topic, respectively.
On initial examination, we observe that Topic~3 contains a lot of words related to the content of didactic designs (\textit{world, literary, water,} etc.). When repeating the process for 4, 5 and 6 clusters, Topic 1 and Topic 2 remain the same, and the rest of the topics split the words accordingly to the content of different didactic designs. In other words, LDA tends to rediscover the various learning activities when the number of clusters is set from 4 to 6. Because we aim to analyze the topics regardless of specific contents, we decided to consider 3 as a sufficient number of clusters to analyze and interpret these first results.

\subsection{Discussion on the exploratory results}

In table~\ref{topwords}, we observe that Topic~1 includes opinion words (\textit{good, interesting, entertaining}) and words that refer to the collaborative learning activity (\textit{work, classmate, activity}). Therefore, we believe that Topic~1 relates to the messages when a learner or a teacher express a subjective attitude about the learning experience or about the work of others. In the Jakobson theory, this kind of messages typically refers to the Emotive function of language (Fig.~2). 
Topic~2 contains greeting words (\textit{hello, hi}), interlocutor markers (\textit{you, boy}) and emoticons. This topic corresponds to the many informal messages written on the platform. Those messages are not directly linked to the didactic activity, they are used to establish and maintain a communication channel (called Phatic function in the Jakobson theory), mainly between students. We also note the significant use of emoticons in this group of words. Emoticons are usually only considered as opinion markers, but we argue that emoticons could also have an important phatic function, in particular among young learners. 
As we said before, Topic~3 regroups words directly related with the content of didactic designs. It also contains a lot of URLs (\textit{http}) indicating that this topic concerns the work itself. This topic is used to share information about a particular context, it is what Jakobson called the Referential function of Language.

Finally, we note that LDA has not identified topics that refers to the the Poetic, Metalingual and Conative functions of Language. Two reasons may explain this: 
(1) There is no message for these functions in our context. It is probably the case of the \emph{Poetic} function. (2) The lexical items are the same as those of an other function. It could be the case of the \emph{Conative} function  and the \emph{Metalingual} function, which can be confused respectively with the \emph{Phatic} function and the \emph{Referential} function.  A more specific study would be necessary to distinguish them.




\begin{table}
\begin{tabular}{l l l}
\textbf{Topic 1 (Emotive)} & \textbf{Topic 2 (Phatic)}& \textbf{Topic 3 (Referential)}\\[0.1cm]
\hline & & \\[-0.2cm]
trabajo \emph{(work)} & XD (EMOTICON) & mundo \emph{(world)} \\
bueno \emph{(good)} & t\'u \emph{(you)} & literario \emph{(literary)} \\
compa\~nero \emph{(classmate)} & ola \emph{(hello)} & texto \emph{(text)} \\
actividad \emph{(activity)} & hola \emph{(hello)} & http (URL) \\
bien \emph{(nice)} & po \emph{(hi)} & cotidiano \emph{(daily)} \\
parecer \emph{(to seem)} & wena \emph{(cool)} & real \emph{(real)} \\
gustar \emph{(to like)} & holi \emph{(hi)} & ser \emph{(to be)} \\
aprender \emph{(to learn)} & :\$ (EMOTICON) &  NUMERIC VALUE\\
interesante \emph{(interesting)} & :p (EMOTICON) & hecho \emph{(done)} \\
entretenido \emph{(entertaining)} & jajaja (\textit{lol}) & vida \emph{(life)} \\
mucho \emph{(very much)} & callar (\emph{to be quiet}) & porque \emph{(because)} \\
terminar \emph{(to finish)} & cabro (\emph{boy}) & onirico \emph{(oneiric)}\\
experiencia \emph{(experience)} & si (\emph{yes}) & historia \emph{(history)}\\
kelluwen \emph{(kelluwen)} & fome (\emph{boring}) & mitico \emph{(mythical)}\\
esperar \emph{(to hope/to wait)} & pap\'a (\emph{dad}) & ejemplo \emph{(example)} \\
trabajar \emph{(to work)} & :) (EMOTICON) & fant\'astico \emph{(fantastic)} \\
nuestro \emph{(our)} & :c (EMOTICON) & agua \emph{(water)} \\
grupo \emph{(group)} & :b (EMOTICON) & pertenecer \emph{(belong to)} \\
subir  \emph{(to climb/to raise)}& :( (EMOTICON) & realidad \emph{(reality)} \\
estar \emph{(to be)} & oli (\emph{hi}) & del \emph{(of the)} \\[0.2cm]
\end{tabular}
\caption{Distribution of the main words (and an approximate translation) in the three topics according to LDA with their associated function of the language (Emotive, Phatic and Referential).}
\label{topwords}
\vspace{-20pt}
\end{table}

\section{Classifying messages}

In the first step, we explored learners' and teachers' messages and identified they might be grouped in 3 discursive categories: Emotive, Phatic and Referential.
The second step consists of evaluating the feasibility of automating the classification of messages. From a reference corpus, we developed and evaluated two statistical models, based on LDA and SVM, to classify messages of the Kelluwen platform.

\subsection{Reference corpus \& Manual annotation}

In order to build a reference corpus, 500 messages from the initial dataset have been randomly selected. Then, 3 human judges determined the main discursive function of each message among the 3 functions identified in the previous step (Emotive, Phatic or Referential). We used the Fleiss' kappa~\cite{Fleiss71} to measure the reliability of agreement between judges. For this task, the Fleiss' Kappa is 0.70 which is a good agreement according to the scale proposed by Landis et al.~\cite{Landis77}. In other words, even if in a few cases this task can be complex for humans (when messages have a really poor syntax or contain only an emoticon for example), most of the time humans are able to understand with what intention a message was transmitted.

Finally, for each message of the reference corpus, the category is the one chosen by at least two judges. There is no cases where all judges disagreed with the two others. The reference corpus contains 240 messages with a Phatic function, 137 messages with an Emotive function and 123 messages with a Referential function.

\subsection{Classification methods}

\subsubsection{SVM}
The first method is based on a SVM supervised learning approach. For each message, we apply the preprocessing as described in the section~3.3 to build a vector representation. In addition to the class of the message, the vector attributes are the generic form of words existing in the entire corpus (without the stop-words) which in this case are 1,543 words. The attribute value is binary and indicates if the word occurs or not in the message. For example, the vector representation of the example (1) contains three attributes with the value "1" : no, entender (\textit{to understand}) and ":-(". Other attributes  (the other 1,540 words) have the value "0". 

\vspace{0.4cm}
\hspace{-0.5cm}(1) No entiendo :-( \hspace{0.4cm} \textit{I don't understand :-(}  \hspace{0.4cm}(\textbf{Emotive})
\vspace{0.4cm}

For our experiment, we used the Sequential minimal optimization (SMO) algorithm available in the Weka library\footnote{http://www.cs.waikato.ac.nz/ml/weka/}. SMO is an algorithm for training SVM, it has the interest to split large SVM learning problems into a series of smaller tasks that can be solved analytically and faster. We trained a SVM model and evaluate it on the reference corpus by using a 10-folds cross-validation technique.


\subsubsection{LDA}
The second method used to automatically classify the messages is based on the LDA approach presented in the section~4. With LDA, each word has a likelihood of belonging to a given topic. Therefore, this classifier simply calculates the likelihood of a message to be in class \emph{$C_i$} as the product of the likelihood of each word of the message to be in class \emph{$C_i$}. The class with the best likelihood is the one kept by this classifier. Because LDA is not directly adapted for supervised classification, this classifier is mainly used as a baseline to evaluate the SVM classifier.

\subsection{Classification results \& Discussion}
\begin{table}
\begin{tabular}{r | r r r | r r r}
& \multicolumn{3}{c|}{\textbf{LDA}} & \multicolumn{3}{c}{\textbf{SVM}}\\
Class & P & R & F1 & P & R & F1 \\[0.1cm]
\hline & & & & & &\\[-0.2cm]
Phatic & 84.9\% & 59.0\% & 69.6\%& \textbf{87.6\%} & \textbf{97.9\%} & \textbf{92.5\%}\\[0.05cm]
Emotive & 74.6\% & 73.0\% &  73.8\% & \textbf{100.0\%} & \textbf{100.0\%} & \textbf{100.0\%}\\[0.05cm]
Referential & 47.2\% & \textbf{76.4}\% & 58.4\% &  \textbf{94.7\%} & 73.2\% & \textbf{82.6\%} \\[0.1cm]
\hline & & & & & &\\[-0.2cm]
Total   & 72.8\% & 67.1\% & 69.8\% & \textbf{92.8\%} & \textbf{92.4\%} &  \textbf{92.6\%} \\[0.1cm]
\end{tabular}
\caption{Precision, Recall and F1-Score nmeasures of learners messages classification into three classes with 2 classifiers: LDA and SVM.}
\label{resultsClassification}
\vspace{-20pt}
\end{table}
To determine the effectiveness of both classifiers, we use 3 classical measures in Text Classification: (1) the \textbf{Precision} (P) measures the system ability to avoid irrelevant solutions, the \textbf{Recall} (R) measures the system ability to provide all the relevant solutions and the \textbf{F1-Score} (F1) is the harmonic mean between Precision and Recall.

\vspace{0.2cm}

\hspace{1cm} P(class$_i$) = $\frac{tp}{tp+fp}$\hspace{1cm} R(class$_i$) = $\frac{tp}{tp+fn}$ 

\vspace{0.2cm}

\hspace{2.5cm}F1(class$_i$) = $\frac{2PR}{P+R}$

\vspace{0.2cm}

Where :
\begin{itemize}
    \item $tp$ (true positive): classifier says "class$_i$" and judge says "class$_i$"
    \item $fp$ (false positive): classifier says "class$_i$" and judge says "NOT class$_i$"
    \item $fn$ (false negative): classifier says "NOT class$_i$" and judge says "class$_i$"
\end{itemize}

A general overview of the results (Table~\ref{resultsClassification}) tends to show that discursive classification of short messages on CSCL platform can be achieved automatically even if texts are noisy. In both methods, we observe that the Emotive class obtain the best F1-Score and is perfectly analyzed by SVM. This result may reflect that young learners do not use a wide range of vocabulary or grammatical constructions when expressing their emotions and that the lexical items are quite specifics to the Emotive class.
In both methods, the Referential messages represented the lowest F1-Score. This result is not surprising because this category regroups a wider diversity of lexical items related to several didactic design contents. The SVM approach tend to forget relevant Referential messages (the Recall is the lowest for this class) and consider that they are Phatic messages (Precision of the Phatic class is the lowest).
One way to improve this aspect could be to collect automatically a set of words related to the didactic design and integrate this knowledge before the classification process.

\section{Evaluating correlations}
In the previous step, we show that a first level of discursive analysis can be achieved automatically on short messages of CSCL platforms with fairly results. In this final step, we investigate how these different types of discourse are correlated or not to other measurable variables.

\subsection{Classification of all the messages}

With the SVM model that we trained and evaluated in the previous step, we classified the 20,128 messages of students from the dataset. According to this automatic classification, the dataset contains 9,984 Phatic messages (median of 61 per classroom group and standard deviation of 122), 3,341 Referential messages (median: 18, standard deviation: 33) and 7,704 Emotive messages (median: 41, standard deviation: 94). We also classified the 1,084 messages from teachers (44 Phatics, 199 Referentials and 841 Emotives).

For each classroom group, we compute the proportion of each of these categories. These variables are called \textit{Phatic}, \textit{Referential} and \textit{Emotive}, with a suffix \textit{\_S} or \textit{\_T} to indicate if it is the proportion in Students' or Teachers' messages.

\subsection{Description of the other variables considered}

For each group, we also consider the following variables: 
\begin{itemize}
\item \textbf{Participation of teachers:} This variable is computed as the percentage of teacher messages over all messages posted in a class group and is named \emph{Teacher Participation}.
\item \textbf{Number of Students:} This represents the number of students that belong to each class group. It is not necessarily the same number of students who participate in posting on the microblog. This factor is named \emph{Students}.
\item \textbf{Progress of the experience:} some class groups never finished the Didactic Design. This variable is computed as the percentage of completed activities in each class group and is called \emph{Progress}.
\item \textbf{Regularity of the activities:} This is measured through obtaining the differences in days between activities, and determining the mean and variance for each class group, so that a small mean and variance implies regularity. The two factors are called \emph{Mean} and \emph{Variance}.
\end{itemize}

\subsection{Correlation results \& Discussion}


\begin{table}
\centering
\tabcolsep=0.09cm
\begin{tabular}{r|rrrrrrrrrr}
& R\_S& P\_S & E\_S & R\_T & P\_T & E\_T & P & S & M & V \\
\hline
Referential\_S & - & & & & & & & & & \\
Phatic\_S & -0.71 & - & & & & & & & & \\
Emotive\_S & 0.10 & -0.77 & - & & & & & & & \\
Referential\_T & 0.11 & 0.04 & -0.16 & - & & & & & & \\
Phatic\_T & -0.19 & 0.09 & 0.04 & -0.41 & - & & & & & \\
Emotive\_T & 0.06 & -0.11 & 0.09 & -0.15 & -0.20 & - & & & & \\
Progress & 0.12 & -0.19 & 0.15 & 0.06 & 0.10 & 0.02 & - & & & \\
Students & 0.00 & 0.09 & -0.14 & 0.06 & -0.03 & 0.15 & 0.23 & - & & \\
Mean & -0.10 & 0.13 & -0.10 & 0.10 & 0.03 & -0.10 & -0.39 & -0.10 & - & \\
Var & 0.04 & -0.02 & -0.01 & -0.05 & 0.05 & -0.02 & -0.06 & -0.06 & 0.64 & -  \\
Teacher Part. & -0.04 & 0.03 & -0.01 & 0.20 & 0.15 & -0.05 & -0.03 & -0.26 & 0.21 & 0.08 \\
\end{tabular}
\caption{Correlation matrix between the discursive attitudes of Students and Teachers and other variables linked to the learning experiences.}
\label{correl}
\vspace{-20pt}
\end{table}

In statistics, the interpretation of a correlation coefficient $x$ ($x \subseteq [-1,+1]$) is a complex task. It is generally considered that there is a strong correlation when $|x|$ is close to 0.7, a weak correlation when $|x|$ is close 0.3 and no correlation when $|x|$ is close to 0. In our case, as we are working with linguistic phenomena, it can be interesting to observe results with less stringent criteria while remaining critical on the interpretation.

\subsubsection{Correlation matrix}

In the correlation matrix (Tab.~\ref{correl}), we observe that there are several strong correlations between the variables Mean/Variance ($x$=0.64), Phatic\_S/Referential\_S ($x$=-0.71) and Emotive\_S/Phatic\_S ($x$=-0.77). These variables are correlated by definition, but we can still observe how the Phatic variable is negatively related to the two other discursive categories.


A more interesting correlation is the one observed between Progress and students' attitudes: Phatic\_S ($x$=-0.19), Referential\_S ($x$=0.12) and Emotive\_S ($x$=0.15). We may think that chatting (excessive Phatic communication) has a bad influence on the success of the learning experience. On the contrary, Emotive or Referential attitudes from students seem to be slightly related to Progress. We may assume that Referential messages tend to be related to a notion of engagement in the learning experience and that Phatic messages, when they are too preponderant, reveal that learners do not focus their attention on the substance of the learning activity. This analysis could provide an interesting way to quickly detect if the teacher must intervene to refocus the activity.

Furthermore, we observe that quantitative participation of the teacher (Teacher Part.) does not seem to be related to Learners' attitudes : Referential\_S ($x$=-0.04), Emotive\_S ($x$=-0.01) and Phatic\_S ($x$=0.03). Nevertheless, if we look more qualitatively at the teachers' attitudes, we can see that Phatic messages from Teachers tend to be negatively related to Referential messages from Students ($x$=-0.19). On the contrary, Emotive messages from Teachers tend to be negatively related to Phatic messages from Students ($x$=-0.11). In other words, some qualitative participation of the teacher may have a relationship with the learners' behaviours. In particular, young learners seem to be sensitive to Emotive feedback from the teacher.



\subsubsection{Principal Components Analysis}

\begin{figure}
\begin{center}
\includegraphics[width = 0.50\textwidth]
{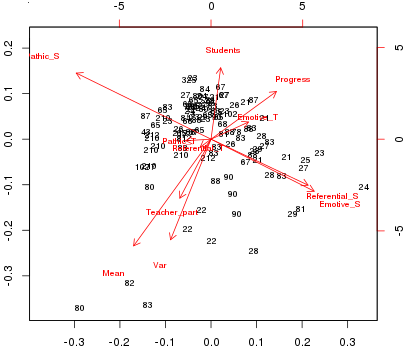}
\caption{PCA analysis over the 11 variables described. The main plane represents 47,41\% of the data variance.}
\label{PCA}
\end{center}
\vspace{-20pt}
\end{figure}

For a graphical analysis, we generated a Principal Component Analysis (PCA) over the data. PCA is a transformation technique for finding patterns in multidimensional data. The Figure~\ref{PCA} shows a two dimensional representation where correlations can be observed as the arrows point to the same (positive correlations), opposite (negative correlations) or orthogonal (no correlations) directions. Each arrow is one of the considered variable, the numbers positioned in the figure correspond to the didactic design ID of each experiences. For example, it shows that the didactic design 21 (Language activity using Youtube) is often related to Progress, Emotive messages from teachers and Referential messages from students. In this sense, the discursive analysis of messages could also be used to help teachers or designers in choosing and developing better learning activities.

Additionally, this figure confirms a quite strong negative correlation between Progress and the average number of days between activities (variable Mean) ($x$=-0.39). We believe that this lack of regularity reflects the teacher's need to be accompanied by indicators to help him manage the pace of the Computer-Supported learning activity over time. A good way to do that could be to monitor the number of Phatic/Emotive/Referential messages over time to tell the teacher when it is time to finish an activity, start a new one or give emotive feedback.
This figure also tends to confirm the correlation between Progress and the proportion of Emotive communication from the teacher. In other words, encouraging emotive feedback from the teacher would be a key aspect to improve learners' engagement and progress in a Computer-Supported learning environment.

\section{Conclusions}

In this article, we show that the analysis of Natural Language interactions between students and teachers provide valuable information that could help teachers' decisions in the context of Computer-Supported Learning. We demonstrated the feasibility to automate the analysis of these interactions even if texts are shorts and noisy.
With a combination of non-supervised and supervised machine learning techniques, we identified that messages of Learners and Teachers can be classified automatically according to three intentions of communication described by Jakobson: create or maintain a communication channel (Phatic), share information about the learning context (Referential) and express an emotion or a subjective attitude (Emotive). 

The results obtained from a correlation calculation bring to light that the types of discourse used by Learners tend to have a relationship with the Teacher qualitative Participation and the Progress of the learning activity. In particular, Emotive communication from the teacher tends to be an interesting aspect to consider to improve learners' engagement in Computer-Supported learning. In this way, the volume of Phatic/Emotive/Referential messages could be monitored automatically over time to provide a complementary indicator of engagement/disengagement. It can also be used to help the teacher evaluate Learning Designs and his own communication.






\bibliography{}

\newpage
a
\end{document}